\documentclass[12pt]{article}
\usepackage{epsf}
\usepackage{graphicx}

\title{ {\bf Asymmetry Parameter of the $K_{1} (1270, 1400)$ by Analyzing the $B\rightarrow K_{1}\nu \overline{\nu}$
Transition  Form Factors within  QCD }}
\author{\vspace{1cm}\\
 M. Bayar \thanks {e-mail:
mbayar@metu.edu.tr}, K. Azizi \thanks {e-mail: e146342@metu.edu.tr}
\\ Physics Department, Middle East Technical University,\\06531
Ankara, Turkey }
 \date{}
\begin{document}
\setlength{\baselineskip}{24pt} \maketitle
\setlength{\baselineskip}{7mm}
\begin{abstract}
Separating the mixture of the $ K_{1}(1270)$ and $K_{1}(1400) $
states, the $B\rightarrow K_{1}(1270, 1400)\nu\overline{\nu}$
transition form factors are calculated in the three-point QCD sum
rules approach. The longitudinal, transverse and total decay widths
as well as the asymmetry parameter, characterizing the polarization
of the axial $K_{1}(1270, 1400)$ and the branching ratio for these
decays are evaluated.
 \end{abstract}
\thispagestyle{empty}
\newpage
\setcounter{page}{1}
\section{Introduction}

The $B\rightarrow K_{1}(1270, 1400)\nu\overline{\nu}$ transitions
are governed by the flavor changing neutral current (FCNC) decay of
$b\rightarrow s\nu\overline{\nu}$ which is of fundamental interest
because of the following reasons: Such transition occurs at loop
level and is forbidden at tree level in the Standard Model (SM).
This transition is a good candidate for searching new physics beyond
the SM and constrains the parameters beyond it. A search for SUSY
particles \cite{Buchalla}, light dark matter \cite{Bird} and also
fourth generation of the quarks  is possible by analyzing such loop
level transition. The $B\rightarrow K_{1}(1270,
1400)\nu\overline{\nu}$ decays also provide a new framework for
precise calculation of the $V_{tb}$ and $V_{ts}$ as elements of the
Cabibbo-Kobayashi-Maskawa (CKM) matrix.

Experimentally, the $K_{1} (1270)$ and $K_{1} (1400)$ are the
mixtures of the strange members of two axial-vector SU(3) octets
$^{3}P_{1}(K_{1}^{A})$ and $^{1}P_{1}(K_{1}^{B})$. The $K_{1} (1270,
1400)$ and $K_{1}^{A,B}$ states are related to each other as
\cite{lee, suzuki}:

\begin{eqnarray}\label{melo}
\mid K_{1}(1270)>&=&\mid K_{1}^{A}>sin \theta+\mid K_{1}^{B}>cos
\theta \nonumber \\
\mid K_{1}(1400)>&=&\mid K_{1}^{A}>cos \theta-\mid K_{1}^{B}>sin
\theta, \nonumber\\
\end{eqnarray}
the angle $\theta$ lies in the interval
$37^{\circ}\leq\theta\leq58^{\circ}$,
$-58^{\circ}\leq\theta\leq-37^{\circ}$ \cite{lee, suzuki,cenk,
burak, cheng}. The sign ambiguity for the mixing angle is related to
the fact that one can add arbitrary phase to the $\mid K_{1}^{A}>$
and $\mid K_{1}^{B}>$. In the recent studies for $B\rightarrow
K_{1}(1270)\gamma$ and $\tau\rightarrow K_{1}(1270)\nu_{\tau}$, the
following values has been obtained for $ \theta$ \cite{hatanaka},
which we are going to use in the present work:
\begin{equation}\label{tetam}
\theta=-(34\pm13)^{\circ}
\end{equation}

 The $B\rightarrow
K_{1}\gamma$ decay has been investigated in the next-to-leading
order in the large energy effective theory (LEET) and in the
framework of light cone sum rules in \cite{Aslam1} and \cite{lee},
respectively. In \cite{Gaur1}, the $B\rightarrow
K_{1}(1270)l\overline{l}$ transition has also been investigated in
the LEET model. In this work, separating the $K_{1} (1270)$ and
$K_{1} (1400)$ states, we analyze the $B\rightarrow K_{1}(1270,
1400)\nu\overline{\nu}$ decay modes in the framework of the
three-point QCD sum rules. First, we calculate the form factors of
the $B$ to axial $\mid K_{1}^{A}>$ and $\mid K_{1}^{B}>$ states.
Then, using the relations among the form factors of the $K_{1}
(1270)$, $K_{1} (1400)$, $\mid K_{1}^{A}>$ and $\mid K_{1}^{B}>$, we
calculate the form factors of the $B\rightarrow K_{1}(1270)$ and
$B\rightarrow K_{1}(1400)$ transitions.

The transition form factors play fundamental role in the evaluating
of the longitudinal, transverse and total decay widths as well as
the asymmetry parameter of $K_{1} (1270, 1400)$. For calculating
these form factors, we use the well established QCD sum rules method
as a non-perturbative method based on the fundamental QCD
lagrangian.

The paper encompasses three sections. In section 2, the form factors
of the $B\rightarrow K_{1}^{A(B)}\nu\overline{\nu}$ transition as
well as the longitudinal and transverse component of the decay width
and asymmetry parameter for $K_{1} (1270, 1400)$ are calculated.
Section 3 is devoted to the numerical analysis and discussions.

\section{Sum rules for the $B\rightarrow K_{1}^{A(B)}\nu\overline{\nu}$ transition form factors }
At the quark level, the $B \rightarrow
K_{1}^{A(B)}\nu\overline{\nu}$ decay proceeds by the loop
$b\rightarrow s\nu\overline{\nu}$ transition. The Hamiltonian
responsible for this transition is given by:
\begin{equation}\label{lelement}
H_{eff}=\frac{G_{F}\alpha_{em}}{2 \sqrt{2}\pi}
C_{10}V_{tb}V_{ts}^{*}~\overline{\nu}
~\gamma_{\mu}(1-\gamma_{5})\nu~\overline{s}
~\gamma_{\mu}(1-\gamma_{5}) b.
\end{equation}
To obtain the transition amplitude for $B\rightarrow
K_{1}^{A(B)}\nu\overline{\nu}$ decay, it is necessary to sandwich
the Eq. (\ref{lelement}) between the initial and final meson states.
\begin{equation}\label{2au}
M=\frac{G_{F}\alpha_{em}}{2 \sqrt{2}\pi}
C_{10}V_{tb}V_{ts}^{*}~\overline{\nu} ~\gamma_{\mu}(1-\gamma_{5})\nu
<K_{1}^{A(B)}(p',\varepsilon)\mid~\overline{s}
~\gamma_{\mu}(1-\gamma_{5}) b\mid B(p)>,
\end{equation}
 where $G_{F}$ is the Fermi constant, $\alpha_{em}$ is the fine structure constant
 at Z mass scale and $V_{ij}$ are the elements of the CKM matrix. Both vector and
 axial vector part of the transition current,
  $~\overline{s}~\gamma_{\mu}(1-\gamma_{5}) b~$, contribute to the
matrix element stated in the Eq. (\ref{2au}). Considering the
Lorentz and parity invariances, this matrix element can be
parameterized in terms of some form factors as follows:
\begin{eqnarray}\label{3au}
<K_{1}^{A(B)}(p',\varepsilon)\mid\overline{s}\gamma_{\mu} b\mid
B(p)>&=&i\left[f_{0}^{A(B)}(q^2)(m_{B}
+m_{K_{1}^{A(B)}})\varepsilon_{\mu}^{\ast}
\right. \nonumber \\
&-&\frac{f_{+}^{A(B)}(q^2)}{(m_{B}+m_{K_{1}^{A(B)}})}(\varepsilon^{\ast}p)P_{\mu}
- \frac{f_-^{A(B)}(q^2)}{(m_{B}+m_{K_{1}^{A(B)}})}(\varepsilon^{\ast}p)q_{\mu}],\nonumber\\
\end{eqnarray}
\begin{eqnarray}\label{4au}
<K_{1}^{A(B)}(p',\varepsilon)\mid\overline{s}\gamma_{\mu}
\gamma_{5}b\mid B(p)> &=&
-\frac{f_{V}^{A(B)}(q^2)}{(m_{B}+m_{K_{1}^{A(B)}})}\varepsilon_{\mu\nu\alpha\beta}
\varepsilon^{\ast\nu}p^\alpha p'^\beta,\end{eqnarray}
 where
$f_{V}^{A(B)}(q^2)$, $f_{0}^{A(B)}(q^2)$, $f_{+}^{A(B)}(q^2)$ and
$f_{-}^{A(B)}(q^2)$ are the transition form factors and
$P_{\mu}=(p+p')_{\mu}$, $q_{\mu}=(p-p')_{\mu}$. Here, we should
mention that the $f_{-}^{A(B)}(q^2)$  form factor does not appear in
the expressions of the decay widths, so we don't consider it in our
calculations. Using the Eqs. ( \ref{melo}, \ref{3au}, \ref{4au}) we
obtain:
\begin{eqnarray}\label{etimelo}
f_{0}^{B\rightarrow
K_{1}(1270)}&=&\frac{m_{B}+m_{K_{1}^{A}}}{m_{B}+m_{K_{1}(1270)}}f_{0}^{B\rightarrow
K_{1}^{A}}sin
\theta+\frac{m_{B}+m_{K_{1}^{B}}}{m_{B}+m_{K_{1}(1270)}}f_{0}^{B\rightarrow
K_{1}^{B}}cos
\theta, \nonumber \\
f_{0}^{B\rightarrow
K_{1}(1400)}&=&\frac{m_{B}+m_{K_{1}^{A}}}{m_{B}+m_{K_{1}(1400)}}f_{0}^{B\rightarrow
K_{1}^{A}}cos
\theta-\frac{m_{B}+m_{K_{1}^{B}}}{m_{B}+m_{K_{1}(1400)}}f_{0}^{B\rightarrow
K_{1}^{B}}sin \theta,
\nonumber \\
\end{eqnarray}
\begin{eqnarray}\label{etimelo1}
f_{+,-,V}^{B\rightarrow
K_{1}(1270)}&=&\frac{m_{B}+m_{K_{1}(1270)}}{m_{B}+m_{K_{1}^{A}}}f_{+,-,V}^{B\rightarrow
K_{1}^{A}}sin
\theta+\frac{m_{B}+m_{K_{1}(1270)}}{m_{B}+m_{K_{1}^{B}}}f_{+,-,V}^{B\rightarrow
K_{1}^{B}}cos
\theta, \nonumber \\
f_{+,-,V}^{B\rightarrow
K_{1}(1400)}&=&\frac{m_{B}+m_{K_{1}(1400)}}{m_{B}+m_{K_{1}^{A}}}f_{+,-,V}^{B\rightarrow
K_{1}^{A}}cos
\theta-\frac{m_{B}+m_{K_{1}(1400)}}{m_{B}+m_{K_{1}^{B}}}f_{+,-,V}^{B\rightarrow
K_{1}^{B}} sin
\theta. \nonumber \\
\end{eqnarray}
For simplicity, we will set $f_{i}^{B\rightarrow
K_{1}^{A(B)}}=f_{i}^{A(B)}$ in the future calculations.

We define the G-parity conserving decay constants of the axial
vector mesons $K_{1}^{A}$ and $K_{1}^{B}$ as
\begin{eqnarray}\label{k1a1}
 <K_{1}^{A}(p',\varepsilon)\mid J_{\nu}=\bar{s}\gamma_{\nu} \gamma_{5}u\mid
0>&=&-if_{K_{1}^{A}}m_{K_{1}^{A}}\varepsilon_{\nu},\nonumber\\
<K_{1}^{B}(p',\varepsilon)\mid J_{\nu\nu'}=\bar{s}\sigma_{\nu\nu'}
\gamma_{5}u\mid
0>&=&f_{K_{1}^{B^{\perp}}}(1~~GeV)(\varepsilon_{\nu}p'_{\nu'}-\varepsilon_{\nu'}p'_{\nu}),
\end{eqnarray}
where $f_{K_{1}^{A}}$ is the scale-independent decay constant of the
$K_{1}^{A}$ meson, however $f_{K_{1}^{B^{\perp}}}$ is the
scale-dependent leptonic constant of the $K_{1}^{B}$ meson. The
$f_{K_{1}^{B^{\perp}}}$ is calculated at scale $1~GeV$. The decay
constants $f_{K_{1}^{A}}$ and $f_{K_{1}^{B^{\perp}}}$ are calculated
in the framework of the light cone QCD sum rules with the help of
the distribution amplitudes of the axial $K_{1}^{A}$ and $K_{1}^{B}$
states in \cite{yangnucl, yangjhep}. On the other hand, the G-party
violating decay constants are defined as
\begin{eqnarray}\label{k1a2}
 <K_{1}^{A}(p',\varepsilon)\mid J_{\nu\nu'}=\bar{s}\sigma_{\nu\nu'}
\gamma_{5}u \mid
0>&=&f_{K_{1}^{A}}~a_{0}^{\perp,K_{1}^{A}}(\varepsilon_{\nu}p'_{\nu'}-\varepsilon_{\nu'}p'_{\nu}),\nonumber\\
<K_{1}^{B}(p',\varepsilon)\mid J_{\nu}=\bar{s}\gamma_{\nu}
\gamma_{5}u \mid
0>&=&if_{K_{1}^{B^{\perp}}}(1~~GeV)~a_{0}^{\parallel,K_{1}^{B}}m_{K_{1}^{B}}~\varepsilon_{\nu},\nonumber\\
\end{eqnarray}
where $a_{0}^{\perp,K_{1}^{A}}$ and $a_{0}^{\parallel,K_{1}^{B}}$
are the zeroth order Gegenbauer moments. They are zero in the
$SU(3)$ symmetry limit. The $a_{0}^{\perp,K_{1}^{A}}$ and
$a_{0}^{\parallel,K_{1}^{B}}$ have been calculated in the framework
of the QCD sum rules \cite{lee, hatanaka, yangnucl}. In these works,
instead of the individual sum rules for the
$a_{0}^{\perp,K_{1}^{A}}$ and $a_{0}^{\parallel,K_{1}^{B}}$, the sum
rules for the combination  of these moments have been obtained.
These calculations led to the relation
$a_{0}^{\perp,K_{1}^{A}}+(0.59\pm0.15)~a_{0}^{\parallel,K_{1}^{B}}=0.17\pm0.11$.
Due to the data for the branching ratio of $B\rightarrow
K_{1}(1270)\gamma$ which is very large than that of the
$B\rightarrow K_{1}(1400)\gamma$ and also $\tau^{-}\rightarrow
K_{1}^{-}(1270)\nu_{\tau}$, the mixing angle and
$a_{0}^{\parallel,K_{1}^{B}}$ should be negative. Assuming that the
G-party violation contribution is about $30~ ^{o}/_{o}$, the values
for $a_{0}^{\perp,K_{1}^{A}}$ and $a_{0}^{\parallel,K_{1}^{B}}$ are
obtained as presented in the numerical analysis section. The Eqs.
(\ref{k1a1}, \ref{k1a2}) show that the main contributions of the
axial $J_{\nu}=\bar{s}\gamma_{\nu} \gamma_{5}u $ and pseudo tensor
$J_{\nu\nu'}=\bar{s}\sigma_{\nu\nu'} \gamma_{5}u$ currents come from
their couplings to the $\mid K_{1}^{A}>$ and $\mid K_{1}^{B}>$,
respectively.

To calculate the form factors, we start considering the following
correlation functions:
\begin{eqnarray}\label{6au}
\Pi _{\mu\nu}^{V;A}(p^2,p'^2,q^2)&=&i^2\int
d^{4}xd^4ye^{-ipx}e^{ip'y}<0\mid T[J _{\nu K_{1}^{A(B)}}(y)
J_{\mu}^{V;A}(0) J_{B}(x)]\mid  0>, \nonumber\\
\Pi_{\mu\nu\nu'}^{V;A}(p^2,p'^2,q^2)&=&i^2\int
d^{4}xd^4ye^{-ipx}e^{ip'y}<0\mid T[J _{\nu\nu' K_{1}^{A(B)}}(y)
J_{\mu}^{V;A}(0) J_{B}(x)]\mid  0>,\nonumber\\
\end{eqnarray}
where $J _{\nu K_{1}^{A(B)}}(y)=\overline{s}\gamma_{\nu}\gamma_{5}
u$ and $J _{\nu\nu' K_{1}^{A(B)}}=\bar{s}\sigma_{\nu\nu'}
\gamma_{5}u$ are the axial vector and pseudo tensor interpolating
currents of the $K_{1}^{A(B)}$ mesons and
$J_{B}(x)=\overline{b}\gamma_{5}u$ is the interpolating current of
$B$ meson. The $J_{\mu}^{V}=~\overline{s}\gamma_{\mu}b $ and
$J_{\mu}^{A}=~\overline{s}\gamma_{\mu}\gamma_{5}b$
 are the vector and axial vector part of the transition currents.
From the general aspect of the QCD sum rules, the above mentioned
correlators are calculated in two different approaches. First, they
are saturated with towers of hadrons with the same quantum numbers
as the interpolating currents called the physical or
phenomenological part and on the other side they describe hadrons as
quarks and gluons interacting with the QCD vacuum called the QCD or
theoretical part. Considering the quark-hadron duality, equating
these two representations of the correlation functions and applying
double Borel transformation with respect to the momentum of the
initial and final states, we get the sum rules for the form factors.
To calculate the correlation functions in the phenomenological side,
we insert complete sets of the intermediate states with the same
quantum numbers as the interpolating currents and sum over the $\mid
K_{1}^{A}>$ and $\mid K_{1}^{B}>$ states. As a result
\begin{eqnarray} \label{7au}
&&\Pi _{\mu\nu}^{V,A}(p^2,p'^2,q^2)=
\nonumber \\
&& \frac{<0\mid J_{K_{1}^{A}}^{\nu} \mid
K_{1}^{A}(p',\varepsilon)><K_{1}^{A}(p',\varepsilon)\mid
J_{\mu}^{V,A}\mid B(p)><B(p)\mid J_{B}\mid
0>}{(p'^2-m_{K_{1}^{A}}^2)(p^2-m_{B}^2)}+
\nonumber \\
&&\frac{<0\mid J_{K_{1}^{B}}^{\nu} \mid
K_{1}^{B}(p',\varepsilon)><K_{1}^{B}(p',\varepsilon)\mid
J_{\mu}^{V,A}\mid B(p)><B(p)\mid J_{B}\mid
0>}{(p'^2-m_{K_{1}^{B}}^2)(p^2-m_{B}^2)}+\ldots
\nonumber \\
\end{eqnarray}
\begin{eqnarray} \label{7au1}
&&\Pi _{\mu\nu\nu'}^{V,A}(p^2,p'^2,q^2)=
\nonumber \\
&& \frac{<0\mid J_{K_{1}^{A}}^{\nu\nu'} \mid
K_{1}^{A}(p',\varepsilon)><K_{1}^{A}(p',\varepsilon)\mid
J_{\mu}^{V,A}\mid B(p)><B(p)\mid J_{B}\mid
0>}{(p'^2-m_{K_{1}^{A}}^2)(p^2-m_{B}^2)}+
\nonumber \\
&&\frac{<0\mid J_{K_{1}^{B}}^{\nu\nu'} \mid
K_{1}^{B}(p',\varepsilon)><K_{1}^{B}(p',\varepsilon)\mid
J_{\mu}^{V,A}\mid B(p)><B(p)\mid J_{B}\mid
0>}{(p'^2-m_{K_{1}^{B}}^2)(p^2-m_{B}^2)}+\ldots
\nonumber \\
\end{eqnarray}
 are obtained. Here, the  $\cdots$ represents contributions
 coming from the higher states and continuum. The vacuum to the hadronic state matrix
 element for $B$ meson in the Eq. (\ref{7au}, \ref{7au1}) are defined in terms of the leptonic decay constant of this
 meson as:
\begin{equation}\label{bbb}
 <B(p)\mid
J_{B}\mid 0>=-i\frac{f_{B}m_{B}^2}{m_{b}+m_{u}}.
\end{equation}
This matrix element for $K_{1}^{A}$ and $K_{1}^{B}$ states are
  presented Eqs. (\ref{k1a1}, \ref{k1a2}).

 Using the above equations and performing summation over the
polarization of the $K_{1}^{A(B)}$ meson we obtain:
\begin{eqnarray}\label{9amplitude}
\Pi_{\mu\nu}^{V}(p^2,p'^2,q^2)&=&i\frac{f_{B}m_{B}^2}{(m_{b}+m_{u})}
\frac{f_{K_{1}^{A}}~m_{K_{1}^{A}}}
{(p'^2-m_{K_{1}^{A}}^2)(p^2-m_{B}^2)}\nonumber\\ &\times&
[f_{0}^{A}g_{\mu\nu} (m_{B}+m_{K_{1}^{A}})
-\frac{f_{+}^{A}P_{\mu}p_{\nu}}{(m_{B}+m_{K_{1}^{A}})} \nonumber
-\frac{f_{-}^{A}q_{\mu}p_{\nu}}{(m_{B}+m_{K_{1}^{A}})}]\\&+&
i\frac{f_{B}m_{B}^2}{(m_{b}+m_{u})}
\frac{f_{K_{1}^{B^{\perp}}}~a_{0}^{\parallel,K_{1}^{B}}~m_{K_{1}^{B}}}
{(p'^2-m_{K_{1}^{B}}^2)(p^2-m_{B}^2)}\nonumber\\ &\times&
[f_{0}^{B}g_{\mu\nu} (m_{B}+m_{K_{1}^{B}})
-\frac{f_{+}^{B}P_{\mu}p_{\nu}}{(m_{B}+m_{K_{1}^{B}})} \nonumber
-\frac{f_{-}^{B}q_{\mu}p_{\nu}}{(m_{B}+m_{K_{1}^{B}})}]\nonumber\\
&+& \mbox{excited states,}\nonumber\\~~
 \Pi_{\mu\nu}^{A}(p^2,p'^2,q^2)&=&
-\varepsilon_{\alpha\beta\mu\nu}p^{\alpha}p'^{\beta}\frac{f_{B}m_{B}^2}{(m_{b}+m_{u})
(m_{B}+m_{K_{1}^{A}})}\frac{f_{K_{1}^{A}}m_{K_{1}^{A}}}
{(p'^2-m_{K_{1}^{A}}^2)(p^2-m_{B}^2)}f_{V}^{A}
\nonumber \\
&-&
\varepsilon_{\alpha\beta\mu\nu}p^{\alpha}p'^{\beta}\frac{f_{B}m_{B}^2}{(m_{b}+m_{u})
(m_{B}+m_{K_{1}^{B}})}\frac{f_{K_{1}^{B^{\perp}}}~a_{0}^{\parallel,K_{1}^{B}}m_{K_{1}^{B}}}
{(p'^2-m_{K_{1}^{B}}^2)(p^2-m_{B}^2)}f_{V}^{B}\nonumber \\&+&
\mbox{excited states.}\nonumber\\
\end{eqnarray}
\begin{eqnarray}\label{9amplitude1}
\Pi_{\mu\nu\nu'}^{V}(p^2,p'^2,q^2)&=&i\frac{f_{B}m_{B}^2}{(m_{b}+m_{u})}
\frac{f_{K_{1}^{B^{\perp}}}}
{(p'^2-m_{K_{1}^{B}}^2)(p^2-m_{B}^2)}\nonumber\\ &\times& [f_{0}^{B}
(m_{B}+m_{K_{1}^{B}})(g_{\nu\mu}p'_{\nu'}-g_{\nu'\mu}p'_{\nu})
+\frac{f_{+}^{B}P_{\mu}}{(m_{B}+m_{K_{1}^{B}})}(p_{\nu'}p'_{\nu}-p_{\nu}p'_{\nu'})\nonumber\\
&+&\frac{f_{-}^{B}q_{\mu}}{(m_{B}+m_{K_{1}^{B}})}(p_{\nu'}p'_{\nu}-p_{\nu}p'_{\nu'})]\nonumber
\\&+&i\frac{f_{B}m_{B}^2}{(m_{b}+m_{u})}
\frac{f_{K_{1}^{A}}~a_{0}^{\perp,K_{1}^{A}}}
{(p'^2-m_{K_{1}^{A}}^2)(p^2-m_{B}^2)}\nonumber\\ &\times& [f_{0}^{A}
(m_{B}+m_{K_{1}^{A}})(g_{\nu\mu}p'_{\nu'}-g_{\nu'\mu}p'_{\nu})
+\frac{f_{+}^{A}P_{\mu}}{(m_{B}+m_{K_{1}^{A}})}(p_{\nu'}p'_{\nu}-p_{\nu}p'_{\nu'})\nonumber\\
&+&\frac{f_{-}^{A}q_{\mu}}{(m_{B}+m_{K_{1}^{A}})}(p_{\nu'}p'_{\nu}-p_{\nu}p'_{\nu'})]
+ \mbox{excited states,}\nonumber\\~~
 \Pi_{\mu\nu\nu'}^{A}(p^2,p'^2,q^2)&=&
(\varepsilon_{\alpha\beta\mu\nu}p'^{\nu'}-\varepsilon_{\alpha\beta\mu\nu'}p'^{\nu})
\frac{p^{\alpha}p'^{\beta}~f_{B}m_{B}^2f_{K_{1}^{B^{\perp}}}f_{V}^{B}}{(m_{b}+m_{u})
(m_{B}+m_{K_{1}^{B}})(p'^2-m_{K_{1}^{B}}^2)(p^2-m_{B}^2)}
\nonumber \\
&+&
(\varepsilon_{\alpha\beta\mu\nu}p'^{\nu'}-\varepsilon_{\alpha\beta\mu\nu'}p'^{\nu})
\frac{p^{\alpha}p'^{\beta}~f_{B}m_{B}^2f_{K_{1}^{A}}~a_{0}^{\perp,K_{1}^{A}}f_{V}^{A}}{(m_{b}+m_{u})
(m_{B}+m_{K_{1}^{A}})(p'^2-m_{K_{1}^{A}}^2)(p^2-m_{B}^2)}\nonumber\\&+&
\mbox{excited states.}
\end{eqnarray}
For extracting the expressions for the form factors
$f_{0}^{A(B)}(q^{2})$ and $f_{+}^{A(B)}(q^{2})$, we choose the
coefficients of the structures $g_{\mu\nu}$ and
$\frac{1}{2}(p_{\mu}p_{\nu}
 + p'_{\mu}p_{\nu})$
from $\Pi_{\mu\nu}^{V} (p^2,p^{\prime 2},q^2)$, respectively and the
structure $i\varepsilon_{\mu\nu\alpha\beta}p'^{\alpha}p^{\beta}$
from $\Pi_{\mu\nu}^{A} (p^2,p^{\prime 2},q^2)$ is considered for the
form factor $f_{V}^{A(B)}(q^{2})$. On the other hand, from the
$\Pi_{\mu\nu\nu'}^{V} (p^2,p^{\prime 2},q^2)$ and
$\Pi_{\mu\nu\nu'}^{A} (p^2,p^{\prime 2},q^2)$  the structures
$i\varepsilon_{\mu\nu'\alpha\beta}p'^{\nu}p^{\alpha}p'^{\beta}$,
$g_{\nu'\mu}p'^{\nu}$, $\frac{1}{2}(p_{\nu'}P_{\mu}p'_{\nu}
 + p_{\nu'}p'_{\nu}q_{\mu})$ are selected for the form factors
 $f_{V}^{A(B)}(q^{2})$, $f_{0}^{A(B)}(q^{2})$ and
 $f_{+}^{A(B)}(q^{2})$, respectively. Here, we stress that there is
 no contribution of the K pole
in the invariant structures  chosen to evaluate the form factors.

Therefore, the correlation functions are written in terms of the
selected structures as:
\begin{eqnarray}\label{kazem}
\Pi_{\mu\nu}^{V}(p,p',q^{2})&=&g_{\mu\nu}\Pi_{0}+\frac{1}{2}(p_{\mu}p_{\nu}
 + p'_{\mu}p_{\nu})\Pi_{+}+....,\nonumber\\
\Pi_{\mu\nu}^{A}(p,p',q^{2})&=&i\varepsilon_{\mu\nu\alpha\beta}p'^{\alpha}p^{\beta}\Pi_{V}+...
\nonumber\\
\Pi_{\mu\nu\nu'}^{V}(p,p',q^{2})&=&g_{\nu'\mu}p'^{\nu}T_{0}+\frac{1}{2}(p_{\nu'}P_{\mu}p'_{\nu}
 + p_{\nu'}p'_{\nu}q_{\mu})T_{+}+....,\nonumber\\
\Pi_{\mu\nu\nu'}^{A}(p,p',q^{2})&=&i\varepsilon_{\mu\nu'\alpha\beta}p'^{\nu}p^{\alpha}p'^{\beta}T_{V}+...
~.\nonumber\\
\end{eqnarray}

The QCD side of the correlation functions are calculated by the help
of the operator product expansion (OPE) in the deep Euclidean
region, where
 $p^2 \ll (m_{b}+m_{u})^2 $ and
$p'^2 \ll (m_{s}+m_{u})^2$. For this aim, we write each
$\Pi_{i}[T_{i}]$  function
 in terms of the perturbative and non-perturbative parts as:
\begin{eqnarray}\label{kazem1}
\Pi_{i}[T_{i}](p,p',q^{2})=\Pi_{i}[T_{i}]^{pert}(p,p',q^{2})
+\Pi_{i}[T_{i}]^{non-pert}(p,p',q^{2}),
\end{eqnarray}
 where  $i$ stands for $0$, $V$ and $+$. The non-perturbative
parts contain  the light quark ($<\bar qq>$) condensates.

 The perturbative parts are written in terms of the double dispersion
 integrals as:
\begin{eqnarray}\label{10au}
\Pi_i[T_{i}]^{pert}=-\frac{1}{(2\pi)^2}\int ds'\int
ds\frac{\rho_{i}[\varrho_{i}](s,s',q^2)}{(s-p^2)(s'-p'^2)}+\textrm{
subtraction terms.}
\end{eqnarray}
The spectral densities $\rho_{i}(s,s',q^2)$ and
$\varrho_{i}(s,s',q^2)$ can be calculated from the usual Feynman
integrals with the help of the Cutkosky rules, i.e., by replacing
the quark propagators with the Dirac-delta functions:
$\frac{1}{p^2-m^2}\rightarrow-2\pi\delta(p^2-m^2),$ implying  all
quarks are real. Calculations lead to the following expressions for
the spectral densities.

\begin{eqnarray}\label{11au}
\rho_{V}(s,s',q^2)&=&4N_{c}I_{0}(s,s',q^2)\left[{(m_{u}-m_{b})A+(m_{u}+m_{s})B}+m_{u}\right],\nonumber\\
\rho_{0}(s,s',q^2)&=&-2N_{c}I_{0}(s,s',q^2)\Bigg[(m_{b}-m_{u})(Au+2Bs'-4C)\nonumber\\
&+&(m_{s}+m_{u})(2As+Bu)\nonumber\\
&+&m_{u}(2m_{b}m_{s}+m_{b}m_{u}-m_{s}m_{u}-m_{u}^{2}-u)\Bigg],\nonumber\\
\nonumber
\rho_{+}(s,s',q^2)&=&2N_{c}I_{0}(s,s',q^2)\Bigg[A(3m_{u}-m_{b})+B(m_{u}+m_{s})\nonumber\\
&+&2(m_{u}-m_{b})(D+E)+mu
\Bigg],\nonumber \\
\nonumber
\end{eqnarray}
\begin{eqnarray}\label{11au1}
\varrho_{V}(s,s',q^2)&=&-8N_{c}I_{0}(s,s',q^2)\left[B+E+F\right],\nonumber\\
\varrho_{0}(s,s',q^2)&=&4N_{c}I_{0}(s,s',q^2)\Bigg[m_{u}(m_{b}-m_{u})-As-2Es-Fu\nonumber\\
&-&B(m_{s}m_{u}+m_{u}^{2}-m_{b}(m_{s}+m_{u})+u)\Bigg],\nonumber\\
\nonumber \varrho_{+}(s,s',q^2)&=&4N_{c}I_{0}(s,s',q^2)\left[B+E+F\right],\nonumber \\
\nonumber
\end{eqnarray}
where
\begin{eqnarray}\label{12}
I_{0}(s,s',q^2)&=&\frac{1}{4\lambda^{1/2}(s,s',q^2)},\nonumber\\
\lambda(a,b,c)&=&a^{2}+b^{2}+c^{2}-2ac-2bc-2ab,\nonumber \\
\Delta&=&m_{b}^{2}-m_{u}^{2}-s,\nonumber \\
\Delta'&=&m_{s}^{2}-m_{u}^{2}-s',\nonumber \\
u&=&s+s'-q^{2},\nonumber \\
A&=&\frac{1}{\lambda(s,s',q^{2})}(\Delta' u-2\Delta s),\nonumber\\
B&=&\frac{1}{\lambda(s,s',q^{2})}(\Delta u-2\Delta' s),\nonumber\\
C&=&\frac{1}{2\lambda(s,s',q^{2})}(\Delta'^{2}s+\Delta^{2} s'-
\Delta\Delta' u+m_{u}^{2}(-4s s'+u^{2}),\nonumber\\
D&=&\frac{1}{\lambda(s,s',q^{2})^{2}}[-6 \Delta\Delta' s'
u+\Delta'^{2}(2 s s'+u^{2})
+2s'(3\Delta^{2} s'+m_{u}^{2}(-4s s'+u^{2}))],\nonumber\\
E&=&\frac{1}{\lambda(s,s',q^{2})^{2}}[-3 \Delta^{2}
s'u+2\Delta\Delta'(2 s s'+u^{2})
-u(3\Delta'^{2} s'+m_{u}^{2}(-4s s'+u^{2}))].\nonumber\\
F&=&\frac{1}{\lambda(s,s',q^{2})^{2}}[6\Delta^{2}
s^2-6\Delta\Delta'su+2m_{u}^{2}s(-4s s'+u^{2})+\Delta^{2}(2s s'+u^{2})].\nonumber\\
 \end{eqnarray}
 The subscripts V, 0 and $+$ correspond to form factors $f_{V}$, $f_{0}$ and $f_{+}$, respectively. In the Eq. (\ref{11au}) $N_{c}=3$ is the number of colors.

 The integration region for the perturbative contribution
 in the Eq. (\ref{10au}) is determined from the condition that the arguments of the
 three $\delta$ functions must vanish simultaneously. The physical
 region in the s and $s'$ plane is described by the following
 non-equality:\\
 \begin{equation}\label{13au}
 -1\leq f(s,s')=\frac{2ss'+(s+s'-q^2)(m_{b}^2-s-m_{u}^2)+(m_{u}^2-m_{s}^2)2s}{\lambda^{1/2}(m_{b}^2,s,m_{u}^2)\lambda^{1/2}(s,s',q^2)}\leq+1.
\end{equation}

  For the contribution of the non-perturbative parts, i.e., the contributions
of the operators with dimensions $d=3$, $4$ and $5$, the following
results are derived:
\begin{eqnarray}\label{14au}
\Pi_{V}^{non-pert}&=&\frac{1}{2}<\overline{q}q>\Bigg[\frac{1}{rr'^{3}}
m_{s}^2(m_{0}^{2}-2m_{u}^2) \nonumber
\\&+&\frac{1}{3r^{2}r'^{2}}[-3m_{u}^2(m_{b}^2+m_{s}^2-q^{2})\nonumber
\\&+&m_{0}^{2}
(m_{b}^2-m_{b}m_{s}+m_{s}^2-q^{2})]\nonumber
\\&-&
\frac{1}{rr'^{2}}m_{s}m_{u}+
\frac{1}{r^{3}r'}m_{b}^{2}(m_{0}^{2}-2m_{u}^{2}) \nonumber
\\&+&\frac{1}{3r^{2}r'}(-2m_{0}^{2}-3m_{b}m_{u}) -\frac{2}{rr'}\Bigg] ,
\nonumber \\
\Pi_{0}^{non-pert}&=&\frac{1}{4}<\overline{q}q>\Bigg[\frac{1}{rr'^{3}}m_{s}^2(m_{0}^{2}-2m_{u}^2)
\nonumber \\
&\times&(m_{b}^{2}-2m_{b}m_{s}+m_{s}^{2}-q^{2})
\nonumber \\
&+& \frac{1}{3r^{2}r'^{2}}(m_{b}^{2}-2m_{b}m_{s}+m_{s}^{2}-q^{2})\nonumber \\
&\times& [-3m_{u}^2(m_{b}^2+m_{s}^2-q^{2})
+m_{0}^{2}(m_{b}^2-m_{b}m_{s}+m_{s}^2-q^{2})]\nonumber\\&+&
\frac{1}{3rr'^{2}}[m_{0}^{2}(m_{b}^2+3m_{b}m_{s}-q^{2})\nonumber\\&-&3m_{u}(m_{s}+m_{u})
(m_{b}^{2}-2m_{b}m_{s}+m_{s}^{2}-q^{2})]
\nonumber\\&+&\frac{1}{r^{3}r'}m_{b}^2(m_{0}^{2}-2m_{u}^{2})
(m_{b}^{2}-2m_{b}m_{s}+m_{s}^{2}-q^{2})\nonumber\\&+&\frac{1}{3r^{2}r'}
[3m_{u}(m_{b}-m_{u})(m_{b}^{2}-2m_{b}m_{s}+m_{s}^{2}-q^{2})\nonumber\\&+&
m_{0}^{2}(3m_{b}m_{s}-m_{s}^2+q^{2})]\nonumber\\&+&\frac{1}{3rr'}(-4m_{0}^{2}-
6m_{b}^{2}-6m_{s}^{2}+3m_{s}m_{u}\nonumber
\\&+&6m_{u}^{2}+m_{b}(4m_{s}+m_{u})+2q^{2})\Bigg],
 \nonumber \\
\Pi_{+}^{non-pert}&=&\frac{1}{4}<\overline{q}q>\Bigg[-\frac{1}{rr'^{3}}
m_{s}^2(m_{0}^{2}-2m_{u}^2)\nonumber \\
&+&\frac{1}{3r^{2}r'^{2}}
[3m_{u}^2(m_{b}^2+m_{s}^2-q^{2})\nonumber \\
&+&m_{0}^{2}(-m_{b}^2+m_{b}m_{s}-m_{s}^2+q^{2})]
\nonumber \\
&+&\frac{1}{rr'^{2}}m_{s}m_{u}
+\frac{1}{4r^{3}r'}m_{b}^2(m_{0}^{2}-2m_{u}^2)
\nonumber \\
&+&\frac{1}{3r^{2}r'}[4m_{0}^{2}-3m_{u}(m_{b}+2m_{u})]
+\frac{2}{rr'}\Bigg],\nonumber \\
\end{eqnarray}

\begin{eqnarray}\label{14au1}
T_{V}^{non-pert}&=&0 ,
\nonumber \\
T_{0}^{non-pert}&=&\frac{1}{2}<\overline{q}q>\Bigg[\frac{1}{rr'^{3}}m_{b}m_{s}^2(m_{0}^{2}-2m_{u}^2)
\nonumber \\
&+& \frac{m_{b}}{3r^2r'^{2}}[-3m_{u}^2(m_{b}^2+m_{s}^2-q^{2})
+m_{0}^{2}(2m_{b}^2-m_{b}m_{s}+2m_{s}^2-2q^{2})]\nonumber\\&+&
\frac{1}{3rr'^{2}}[-3m_{u}(-m_{b}^2+m_{b}m_{s}-m_{s}^2+m_{s}m_{u}-q^{2})
+m_{0}^{2}(m_{b}+2m_{s})]
\nonumber\\&+&\frac{1}{r^{3}r'}m_{b}^3(m_{0}^{2}-2m_{u}^{2})
\nonumber\\&-&\frac{m_{b}}{3r^{2}r'}
[3m_{u}(m_{b}-m_{u})+m_{0}^{2}]+ \frac{2m_{b}}{rr'}\Bigg],
 \nonumber \\
T_{+}^{non-pert}&=&-\frac{1}{2}<\overline{q}q>[\frac{m_{b}m_{0}^{2}}{3r^2r'^{2}}
+\frac{m_{u}}{rr'^{2}}],
\end{eqnarray}
where $r=p^{2}-m_{b}^{2}$ and $r'=p'^{2}-m_{c}^{2}$.

 Equating the phenomenological expression
 given in the Eqs. (\ref{9amplitude},\ref{9amplitude1}) and the
OPE expression given by Eqs. (\ref{kazem1}-\ref{14au1}), and
applying double Borel transformations with respect to the variables
$p^2$ and $p'^2$ ($p^2\rightarrow M_{1}^2,~p'^2\rightarrow M_{2}^2$)
in order to suppress the contributions of the higher states and
continuum, the QCD sum rules for the combinations of the form
factors $f_{V}^{A(B)}$, $f_{0}^{A(B)}$ and $f_{+}^{A(B)}$ are
obtained:
\begin{eqnarray}\label{15auset1}
&&\frac{f_{K_{1}^{A}}m_{K_{1}^{A}}}{(m_{B}+m_{K_{1}^{A}})}e^{-m^2_{K_{1}^{A}}/M_{2}^2}f^{A}_{V}(q^{2})
+\frac{f_{K_{1}^{B^{\perp}}}~a_{0}^{\parallel,K_{1}^{B}}m_{K_{1}^{B}}}{(m_{B}+m_{K_{1}^{B}})}e^{-m^2_{K_{1}^{B}}/M_{2}^2}f^{B}_{V}(q^{2})\nonumber\\
&=&-\frac{(m_{b}+m_{u})}{f_{B}m_{B}^2}e^{m_{B}^2/M_{1}^2}
\Bigg\{-\frac{1}{(2\pi)^2}\int_{(m_{s}+m_{u})^{2}}^{s_0'} ds'
\int_{(m_{b}+m_{u})^{2}}^{s_0}
ds~\rho_{V}(s,s',q^2)\nonumber\\&\times&\theta[1-f^{2}(s,s')]e^{-s/M_{1}^2-s'/M_{2}^2}+\hat{B}(\Pi_{V}^{non-pert})\Bigg\},\nonumber\\\nonumber\\
&&\frac{f_{K_{1}^{A}}~a_{0}^{\perp,K_{1}^{A}}}{(m_{B}+m_{K_{1}^{A}})}e^{-m^2_{K_{1}^{A}}/M_{2}^2}f^{A}_{V}(q^{2})
+\frac{f_{K_{1}^{B^{\perp}}}}{(m_{B}+m_{K_{1}^{B}})}e^{-m^2_{K_{1}^{B}}/M_{2}^2}f^{B}_{V}(q^{2})\nonumber\\
&=&-\frac{(m_{b}+m_{u})}{f_{B}m_{B}^2}e^{m_{B}^2/M_{1}^2}
\Bigg\{-\frac{1}{(2\pi)^2}\int_{(m_{s}+m_{u})^{2}}^{s_0'} ds'
\int_{(m_{b}+m_{u})^{2}}^{s_0}
ds~\varrho_{V}(s,s',q^2)\nonumber\\&\times&\theta[1-f^{2}(s,s')]e^{-s/M_{1}^2-s'/M_{2}^2}+\hat{B}(T_{V}^{non-pert})\Bigg\},\nonumber\\
\end{eqnarray}
\begin{eqnarray}\label{15auset2}
&&f_{K_{1}^{A}}m_{K_{1}^{A}}(m_{B}+m_{K_{1}^{A}})e^{-m^2_{K_{1}^{A}}/M_{2}^2}f^{A}_{0}(q^{2})
+f_{K_{1}^{B^{\perp}}}~a_{0}^{\parallel,K_{1}^{B}}m_{K_{1}^{B}}(m_{B}+m_{K_{1}^{B}})e^{-m^2_{K_{1}^{B}}/M_{2}^2}f^{B}_{0}(q^{2})\nonumber\\
&=&\frac{(m_{b}+m_{u})}{f_{B}m_{B}^2}e^{m_{B}^2/M_{1}^2}
\Bigg\{-\frac{1}{(2\pi)^2}\int_{(m_{s}+m_{u})^{2}}^{s_0'} ds'
\int_{(m_{b}+m_{u})^{2}}^{s_0}
ds~\rho_{0}(s,s',q^2)\nonumber\\&\times&\theta[1-f^{2}(s,s')]e^{-s/M_{1}^2-s'/M_{2}^2}+\hat{B}(\Pi_{0}^{non-pert})\Bigg\},\nonumber\\\nonumber\\
&&-f_{K_{1}^{A}}~a_{0}^{\perp,K_{1}^{A}}(m_{B}+m_{K_{1}^{A}})e^{-m^2_{K_{1}^{A}}/M_{2}^2}f^{A}_{0}(q^{2})
-f_{K_{1}^{B^{\perp}}}(m_{B}+m_{K_{1}^{B}})e^{-m^2_{K_{1}^{B}}/M_{2}^2}f^{B}_{0}(q^{2})\nonumber\\
&=&\frac{(m_{b}+m_{u})}{f_{B}m_{B}^2}e^{m_{B}^2/M_{1}^2}
\Bigg\{-\frac{1}{(2\pi)^2}\int_{(m_{s}+m_{u})^{2}}^{s_0'} ds'
\int_{(m_{b}+m_{u})^{2}}^{s_0}
ds~\varrho_{0}(s,s',q^2)\nonumber\\&\times&\theta[1-f^{2}(s,s')]e^{-s/M_{1}^2-s'/M_{2}^2}+\hat{B}(T_{0}^{non-pert})\Bigg\},\nonumber\\
\end{eqnarray}
\begin{eqnarray}\label{15auset3}
&&-\frac{f_{K_{1}^{A}}m_{K_{1}^{A}}}{(m_{B}+m_{K_{1}^{A}})}e^{-m^2_{K_{1}^{A}}/M_{2}^2}f^{A}_{+}(q^{2})
-\frac{f_{K_{1}^{B^{\perp}}}~a_{0}^{\parallel,K_{1}^{B}}m_{K_{1}^{B}}}{(m_{B}+m_{K_{1}^{B}})}e^{-m^2_{K_{1}^{B}}/M_{2}^2}f^{B}_{+}(q^{2})\nonumber\\
&=&\frac{(m_{b}+m_{u})}{f_{B}m_{B}^2}e^{m_{B}^2/M_{1}^2}
\Bigg\{-\frac{1}{(2\pi)^2}\int_{(m_{s}+m_{u})^{2}}^{s_0'} ds'
\int_{(m_{b}+m_{u})^{2}}^{s_0}
ds~\rho_{+}(s,s',q^2)\nonumber\\&\times&\theta[1-f^{2}(s,s')]e^{-s/M_{1}^2-s'/M_{2}^2}+\hat{B}(\Pi_{+}^{non-pert})\Bigg\},\nonumber\\\nonumber\\
&&\frac{f_{K_{1}^{A}}~a_{0}^{\perp,K_{1}^{A}}}{(m_{B}+m_{K_{1}^{A}})}e^{-m^2_{K_{1}^{A}}/M_{2}^2}f^{A}_{+}(q^{2})
+\frac{f_{K_{1}^{B^{\perp}}}}{(m_{B}+m_{K_{1}^{B}})}e^{-m^2_{K_{1}^{B}}/M_{2}^2}f^{B}_{+}(q^{2})\nonumber\\
&=&\frac{(m_{b}+m_{u})}{f_{B}m_{B}^2}e^{m_{B}^2/M_{1}^2}
\Bigg\{-\frac{1}{(2\pi)^2}\int_{(m_{s}+m_{u})^{2}}^{s_0'} ds'
\int_{(m_{b}+m_{u})^{2}}^{s_0}
ds~\varrho_{+}(s,s',q^2)\nonumber\\
&\times&\theta[1-f^{2}(s,s')]e^{-s/M_{1}^2-s'/M_{2}^2}
+\hat{B}(T_{+}^{non-pert})\Bigg\}.
\end{eqnarray}

In each set of the above equations, we have two equations with two
unknowns (form factors). To obtain the form factors
$f_{V}^{A}(q^{2})$, $f_{V}^{B}(q^{2})$, $f_{0}^{A}(q^{2})$,
$f_{0}^{B}(q^{2})$, $f_{+}^{A}(q^{2})$ and $f_{+}^{B}(q^{2})$, we
solve each set simultaneously. Finally, we obtain the sum rules as
following:

\begin{eqnarray}\label{fva}
f^{A}_{V}(q^{2})&=&-\frac{1}{12f_{B}f_{K_{1}^{A}}m_{B}^{2}(m_{K_{1}^{A}}
-a_{0}^{\perp,K_{1}^{A}}a_{0}^{\parallel,K_{1}^{B}}m_{K_{1}^{B}})\pi^{2}}
\nonumber\\&\times&\Bigg\{e^{m_{K_{1}^{A}}^2/M_{2}^2+m_{B}^2/M_{1}^2}
(m_{B}+m_{K_{1}^{A}})(m_{b}+m_{u})\Bigg[4\pi^{2}\hat{B}(\Pi_{V}^{non-pert})\nonumber\\
&+&3a_{0}^{\parallel,K_{1}^{B}}m_{K_{1}^{B}}\int_{(m_{s}+m_{u})^{2}}^{s_0'}
ds' \int_{(m_{b}+m_{u})^{2}}^{s_0}
ds~\varrho_{V}(s,s',q^2)\theta[1-f^{2}(s,s')]e^{-s/M_{1}^2-s'/M_{2}^2}\nonumber\\
&-&3\int_{(m_{s}+m_{u})^{2}}^{s_0'} ds'
\int_{(m_{b}+m_{u})^{2}}^{s_0}
ds~\rho_{V}(s,s',q^2)\theta[1-f^{2}(s,s')]e^{-s/M_{1}^2-s'/M_{2}^2}\Bigg]\Bigg\},\nonumber\\
\end{eqnarray}
\begin{eqnarray}\label{fvb}
f^{B}_{V}(q^{2})&=&\frac{1}{12f_{B}f_{K_{1}^{B^{\perp}}}m_{B}^{2}(m_{K_{1}^{A}}-a_{0}^{\perp,K_{1}^{A}}a_{0}^{\parallel,K_{1}^{B}}m_{K_{1}^{B}})\pi^{2}}
\nonumber\\&\times&\Bigg\{e^{m_{K_{1}^{A}}^2/M_{2}^2+m_{B}^2/M_{1}^2}(m_{B}+m_{K_{1}^{B}})(m_{b}+m_{u})
\Bigg[4\pi^{2}a_{0}^{\perp,K_{1}^{A}}\hat{B}(\Pi_{V}^{non-pert})\nonumber\\
&+&3m_{K_{1}^{A}}\int_{(m_{s}+m_{u})^{2}}^{s_0'} ds'
\int_{(m_{b}+m_{u})^{2}}^{s_0}
ds~\varrho_{V}(s,s',q^2)\theta[1-f^{2}(s,s')]e^{-s/M_{1}^2-s'/M_{2}^2}\nonumber\\
&-&3a_{0}^{\perp,K_{1}^{A}}\int_{(m_{s}+m_{u})^{2}}^{s_0'} ds'
\int_{(m_{b}+m_{u})^{2}}^{s_0}
ds~\rho_{V}(s,s',q^2)\theta[1-f^{2}(s,s')]e^{-s/M_{1}^2-s'/M_{2}^2}\Bigg]\Bigg\},\nonumber\\
\end{eqnarray}

\begin{eqnarray}\label{foa}
f^{A}_{0}(q^{2})&=&\frac{1}{12f_{B}f_{K_{1}^{A}}m_{B}^{2}(m_{B}+m_{K_{1}^{A}})(m_{K_{1}^{A}}-a_{0}^{\perp,K_{1}^{A}}a_{0}^{\parallel,K_{1}^{B}}m_{K_{1}^{B}})\pi^{2}}
\nonumber\\&\times&\Bigg\{e^{m_{K_{1}^{A}}^2/M_{2}^2+m_{B}^2/M_{1}^2}(m_{b}+m_{u})\Bigg[4\pi^{2}\Bigg(\hat{B}(\Pi_{0}^{non-pert})
-a_{0}^{\parallel,K_{1}^{B}}m_{K_{1}^{B}}\hat{B}(T_{0}^{non-pert})\Bigg)\nonumber\\
&-&3\int_{(m_{s}+m_{u})^{2}}^{s_0'} ds'
\int_{(m_{b}+m_{u})^{2}}^{s_0}
ds~\rho_{0}(s,s',q^2)\theta[1-f^{2}(s,s')]e^{-s/M_{1}^2-s'/M_{2}^2}\nonumber\\
&-&3a_{0}^{\parallel,K_{1}^{B}}m_{K_{1}^{B}}\int_{(m_{s}+m_{u})^{2}}^{s_0'}
ds' \int_{(m_{b}+m_{u})^{2}}^{s_0}
ds~\varrho_{0}(s,s',q^2)\theta[1-f^{2}(s,s')]e^{-s/M_{1}^2-s'/M_{2}^2}\Bigg]\Bigg\},\nonumber\\
\end{eqnarray}
\begin{eqnarray}\label{fob}
f^{B}_{0}(q^{2})&=&-\frac{1}{12f_{B}f_{K_{1}^{B^{\perp}}}m_{B}^{2}(m_{B}+m_{K_{1}^{B}})(-m_{K_{1}^{A}}+a_{0}^{\perp,K_{1}^{A}}a_{0}^{\parallel,K_{1}^{B}}m_{K_{1}^{B}})\pi^{2}}
\nonumber\\&\times&\Bigg\{e^{m_{K_{1}^{A}}^2/M_{2}^2+m_{B}^2/M_{1}^2}(m_{b}+m_{u})\Bigg[4\pi^{2}\Bigg(-a_{0}^{\perp,K_{1}^{A}}\hat{B}(\Pi_{+}^{non-pert})
+m_{K_{1}^{A}}\hat{B}(T_{+}^{non-pert})\Bigg)\nonumber\\
&+&3a_{0}^{\perp,K_{1}^{A}}\int_{(m_{s}+m_{u})^{2}}^{s_0'} ds'
\int_{(m_{b}+m_{u})^{2}}^{s_0}
ds~\rho_{0}(s,s',q^2)\theta[1-f^{2}(s,s')]e^{-s/M_{1}^2-s'/M_{2}^2}\nonumber\\
&+&3m_{K_{1}^{A}}\int_{(m_{s}+m_{u})^{2}}^{s_0'} ds'
\int_{(m_{b}+m_{u})^{2}}^{s_0}
ds~\varrho_{0}(s,s',q^2)\theta[1-f^{2}(s,s')]e^{-s/M_{1}^2-s'/M_{2}^2}\Bigg]\Bigg\},\nonumber\\
\end{eqnarray}

\begin{eqnarray}\label{fpa}
f^{A}_{+}(q^{2})&=&\frac{1}{12f_{B}f_{K_{1}^{A}}m_{B}^{2}(m_{K_{1}^{A}}-a_{0}^{\perp,K_{1}^{A}}a_{0}^{\parallel,K_{1}^{B}}m_{K_{1}^{B}})\pi^{2}}
\nonumber\\&\times&\Bigg\{e^{m_{K_{1}^{A}}^2/M_{2}^2+m_{B}^2/M_{1}^2}(m_{B}+m_{K_{1}^{A}})(m_{b}+m_{u})\nonumber\\&\times&\Bigg[-4\pi^{2}
\Bigg(\hat{B}(\Pi_{+}^{non-pert})
-a_{0}^{\parallel,K_{1}^{B}}m_{K_{1}^{B}}\hat{B}(T_{+}^{non-pert})\Bigg)\nonumber\\
&+&3\int_{(m_{s}+m_{u})^{2}}^{s_0'} ds'
\int_{(m_{b}+m_{u})^{2}}^{s_0}
ds~\rho_{+}(s,s',q^2)\theta[1-f^{2}(s,s')]e^{-s/M_{1}^2-s'/M_{2}^2}\nonumber\\
&+&3a_{0}^{\parallel,K_{1}^{B}}m_{K_{1}^{B}}\int_{(m_{s}+m_{u})^{2}}^{s_0'}
ds' \int_{(m_{b}+m_{u})^{2}}^{s_0}
ds~\varrho_{+}(s,s',q^2)\theta[1-f^{2}(s,s')]e^{-s/M_{1}^2-s'/M_{2}^2}\Bigg]\Bigg\},\nonumber\\
\end{eqnarray}

\begin{eqnarray}\label{fpb}
f^{B}_{+}(q^{2})&=&-\frac{1}{12f_{B}f_{K_{1}^{B^{\perp}}}m_{B}^{2}(m_{K_{1}^{A}}-a_{0}^{\perp,K_{1}^{A}}a_{0}^{\parallel,K_{1}^{B}}m_{K_{1}^{B}})\pi^{2}}
\nonumber\\&\times&\Bigg\{e^{m_{K_{1}^{A}}^2/M_{2}^2+m_{B}^2/M_{1}^2}(m_{B}+m_{K_{1}^{B}})(m_{b}+m_{u})\nonumber\\&\times&
\Bigg[4\pi^{2}\Bigg(-a_{0}^{\perp,K_{1}^{A}}\hat{B}(\Pi_{+}^{non-pert})
+m_{K_{1}^{A}}\hat{B}(T_{+}^{non-pert})\Bigg)\nonumber\\
&+&3a_{0}^{\perp,K_{1}^{A}}\int_{(m_{s}+m_{u})^{2}}^{s_0'} ds'
\int_{(m_{b}+m_{u})^{2}}^{s_0}
ds~\rho_{+}(s,s',q^2)\theta[1-f^{2}(s,s')]e^{-s/M_{1}^2-s'/M_{2}^2}\nonumber\\
&+&3m_{K_{1}^{A}}\int_{(m_{s}+m_{u})^{2}}^{s_0'} ds'
\int_{(m_{b}+m_{u})^{2}}^{s_0}
ds~\varrho_{+}(s,s',q^2)\theta[1-f^{2}(s,s')]e^{-s/M_{1}^2-s'/M_{2}^2}\Bigg]\Bigg\}.\nonumber\\
\end{eqnarray}

In the above equation, in order to subtract the contributions of the
higher states and the continuum, the quark-hadron duality assumption
is used, i.e.,
\begin{eqnarray}
\rho^{higher states}(s,s') = \rho^{OPE}(s,s') \theta(s-s_0)
\theta(s'-s'_0).
\end{eqnarray}
$~~~~~~~~~~~~~~~~~~~~~~~~~~~~~~~~~~~~~~~~~~~~~~~~~~~~~$

 At the end of this section, we would like to calculate the longitudinal and
  transverse component of the differential decay width in terms of the form factors of $K_{1} (1270)$ and $K_{1} (1400)$ (see  Eq. (\ref{etimelo})
  for the relation between the $K_{1}^{A(B)}$ and $K_{1} (1270, 1400)$ form factors). After some calculations, we obtain the longitudinal and transverse
components of the differential decay width as
\begin{eqnarray}\label{28au}
\frac{d\Gamma_{L}}{dq^2}&=&\frac{3G_{F}^{2}\alpha_{em}^{2}
\mid\overrightarrow{p'}\mid}{192\sqrt{2}\pi^{5}} \mid
V_{tb}V_{ts}^{*}\mid^{2} \mid C_{10}\mid^{2} \Bigg\{\frac{
\mid\overrightarrow{p'}\mid^{2}(m_{K_{1}}^{2}-m_{B}^{2}+q^{2})
}{m_{K_{1}}^{2}}Re[f_{0}f_{+}]\nonumber\\&+&\frac{(m_{B}+m_{K_{1}})^{2}
(m_{B}^{2}\mid\overrightarrow{p'}\mid^{2}+m_{K_{1}}^{2}q^{2})}{2m_{B}^{2}m_{K_{1}}^{2}}
\mid
f_{0}\mid^{2}\nonumber\\&+&\frac{1}{8m_{B}^{2}m_{K_{1}}^{2}(m_{B}+m_{K_{1}})^{2}}\Bigg[
m_{K_{1}}^{2}q^{6}\mid
f_{V}\mid^{2}\nonumber\\&+&4m_{B}^{2}((m_{B}-m_{K_{1}})^{2}-q^{2})((m_{B}+m_{K_{1}})^{2}-q^{2})
\mid\overrightarrow{p'}\mid^{2}\mid
f_{+}\mid^{2}\Bigg]\nonumber\\&+&\frac{1}{8m_{B}^{2}(m_{B}+m_{K_{1}})^{2}}\Bigg[
m_{B}^{4}+m_{K_{1}}^{4}-2m_{K_{1}}^{2}q^{2}\nonumber\\&-&2m_{B}^{2}(m_{K_{1}}^{2}
+2\mid\overrightarrow{p'}\mid^{2}+q^{2})\Bigg]\mid
f_{V}\mid^{2}q^{2}\Bigg\},
\nonumber \\
\end{eqnarray}
\begin{eqnarray}\label{2801au}
\frac{d\Gamma_{T}}{dq^2}&=&\frac{3G_{F}^{2}\alpha_{em}^{2}
\mid\overrightarrow{p'}\mid}{192\sqrt{2}\pi^{5}} \mid
V_{tb}V_{ts}^{*}\mid^{2} \mid C_{10}\mid^{2}q^{2}
\Bigg\{\frac{(m_{B}+m_{K_{1}})^{2}}{2m_{B}^{2}} \mid
f_{0}\mid^{2}\nonumber\\&+&\frac{((m_{B}-m_{K_{1}})^{2}-q^{2})((m_{B}+m_{K_{1}})^{2}-q^{2})}
{8m_{B}^{2}(m_{B}+m_{K_{1}})^{2}}\mid f_{V}\mid^{2}\Bigg\}.
\nonumber \\
\end{eqnarray}
The total decay width and the asymmetry parameter $\alpha$,
characterizing the polarization of the $K_{1}$ meson are define as:
\begin{eqnarray}\label{total}
 \frac{d\Gamma_{tot}}{dq^2}&=&\frac{d\Gamma_{L}}{dq^2}+2\frac{d\Gamma_{T}}{dq^2},\nonumber\\
 \alpha&=&2\frac{d\Gamma_{L}}{dq^2}/\frac{d\Gamma_{T}}{dq^2}-1.
\end{eqnarray}

\section{Numerical analysis}
In this section, we present our numerical analysis of the form
factors $f_{V}~,~f_{0}~$ and $f_{+}$, longitudinal, transverse and
total decay width, branching ratio and the asymmetry parameter
$\alpha$, characterizing the polarization of the $K_{1}$ meson. The
sum rules expressions for the form factors and also the expression
for the decay widths depict that the main input parameters entering
the expressions are the Wilson coefficient $C_{10}$, elements of the
CKM matrix $V_{tb}$ and $V_{ts}^{*}$, the leptonic decay constants;
$f_{B}$ and $f_{K_{1}^{A(B^{\perp})}}$, the Borel parameters
$M_{1}^2$ and $M_{2}^2$, as well as the continuum thresholds $s_{0}$
and $s'_{0}$. In further numerical analysis, we choose the values of
the leptonic decay constants, the CKM matrix elements, the Wilson
coefficient $C_{10}$, the quark and meson masses  as:
$C_{10}=-4.669$
 \cite{Buras,Bashiry},  $\mid
V_{tb}\mid=0.77^{+0.18}_{-0.24}$, $\mid
V_{ts}\mid=(40.6\pm2.7)\times10^{-3}$ \cite {Ceccucci},
$f_{K_{1}^{A}} =(250\pm13)  ~MeV $, $f_{K_{1}^{B^{\perp}}}
=(190\pm10) ~MeV $ \cite{lee, hatanaka},
  $f_{B}=0.14\pm 0.01 ~GeV$,
\cite{13}, $ m_{c}= 1.25\pm0.09~ GeV$, $m_{s}=95\pm25 ~MeV$, $m_{b}
=(4.7\pm0.07)~GeV$, $m_{d}=(3-7)~MeV$, $m_{B}=5.279~GeV$,  $
m_{K_{1}}(1270)=1.27~GeV$, $ m_{K_{1}}(1400)=1.40~GeV$\cite{Yao},
$m_{K_{1}^{A}} =(1.31\pm0.06)  ~GeV $, $m_{K_{1}^{B}} =(1.34\pm0.08)
~MeV $ \cite{lee, hatanaka,onalti}. The zeroth order Gegenbaurer
moments are taken to be $a_{0}^{\parallel,K_{1}^{B}}(1
~GeV)=-0.19\pm0.07 $ and $a_{0}^{\perp,K_{1}^{A}}(1
~GeV)=0.27^{+0.03} _{-0.17}$ \cite{onalti}.

The expressions for the form factors  contain also four auxiliary
parameters: Continuum thresholds $s_{0}$  and $s'_{0}$ and Borel
mass squares $M_{1}^2$ and $M_{2}^2$. These are not physical
quantities, hence the physical quantities, form factors, should be
independent of them. The continuum thresholds $s_0$ and $s_0'$ in
the $B$ and $K_{1}$ channels are
 determined from the conditions that
guarantee the sum rules to have the best stability in the allowed
$M_1^2$ and $M_2^2$ region. The values of continuum thresholds
calculated from the two--point QCD sum rules are taken to be
$s_0=(35\pm5)~GeV^2$ and $s_0^\prime=(4\pm1)~GeV^2$.  The working
regions for $M_1^2$ and $M_2^2$ are determined by requiring that not
only contributions of the   higher states and continuum are
effectively suppressed, but the contributions of the higher
dimensional operators are small. Both conditions are satisfied in
the  regions $10~GeV^2 \le M_1^2 \le 22~GeV^2$ and $3~GeV^2 \le
M_2^2 \le 8~GeV^2$. The value of the form factors at $q^2=0$  are
given in Table 1.
\begin{table}[h]
\centering
\begin{tabular}{|c||c c c|} \hline
~~ & $f_{V}(0)$ & $f_{0}(0)$ & $f_{+}(0)$
\\\cline{1-4}\hline\hline
$B\rightarrow K_{1}(1270)\nu\overline{\nu}$  & $0.57\pm0.21$ &
$0.24\pm0.10$ & $0.39\pm0.14$
\\\cline{1-4}\hline
$B\rightarrow K_{1}(1400)\nu\overline{\nu}$  & $0.40\pm0.15$ &
$0.17\pm0.07$ & $0.29\pm0.10$
\\\cline{1-4}\hline
\end{tabular}
\vspace{0.8cm} \caption{The value of the form factors at $q^2=0$,
$M_{1}^{2}=15~GeV^2$, $M_{2}^{2}=4~GeV^2$.} \label{tab:1}
\end{table}
\begin{table}[h] \centering
\begin{tabular}{|c||c|c|c|c|c|c|}\hline
  & $f_{V}^{K_{1}(1270)}$  & $f_{0}^{K_{1}(1270)}$ & $f_{+}^{K_{1}(1270)}$
  & $f_{V}^{K_{1}(1400)}$  & $f_{0}^{K_{1}(1400)}$ & $f_{+}^{K_{1}(1400)}$\\\cline{1-7} \hline \hline
 a  & -5.70 & -3.05 & -4.05 & -4.05 & -0.18 & -0.48\\\cline{1-7}
 \hline
  b  & 6.28 & 3.29 & 4.43 & 4.43 & 0.35 & 0.75\\\cline{1-7}
 \hline
  $m_{fit} $ & 6.99 & 9.41 & 6.64 & 6.63 & 6.47 & 6.16\\\cline{1-7}
 \hline
  \end{tabular}
 \vspace{0.8cm}
\caption{Parameters appearing in the fit function for form factors
of the $B\rightarrow K_{1}(1270, 1400)\overline{\nu}\nu$ at
$M_{1}^{2}=15~GeV^2$, $M_{2}^{2}=4~GeV^2$.} \label{tab:2}
\end{table}

 The sum rules expressions for the form factors are
truncated at $12 ~GeV^2$, about  $4 ~GeV^2$ below the upper limit of
the $q^{2}$ which is about $16 ~(15)~GeV^2$ for $K_{1}(1270)$
($K_{1}(1400)$). In order to extend our results to the whole
physical region, i.e., $0\leq q^{2}<(m_{B}-m_{K_{1}})^{2}$ and for
the reliability of the sum rules in the full physical region, we
look for a fit parametrization such that in the region $0\leq q^{2}
\leq 7~GeV^{2}$, these parameterizations coincide with the sum rules
predictions. To find the extrapolation of the form factors, we
choose the following fit function
 \begin{equation}\label{17au}
 f_{i}(q^2)=\frac{a}{(1-\frac{q^{2}}{m_{fit}^{2}})}+\frac{b}{(1-\frac{q^{2}}{m_{fit}^{2}})^{2}}.
\end{equation}
 The values for a, b and
$m_{fit}$ are given in Table 2.

\begin{figure}
\vspace*{-1cm}
\begin{center}
\includegraphics[width=10cm]{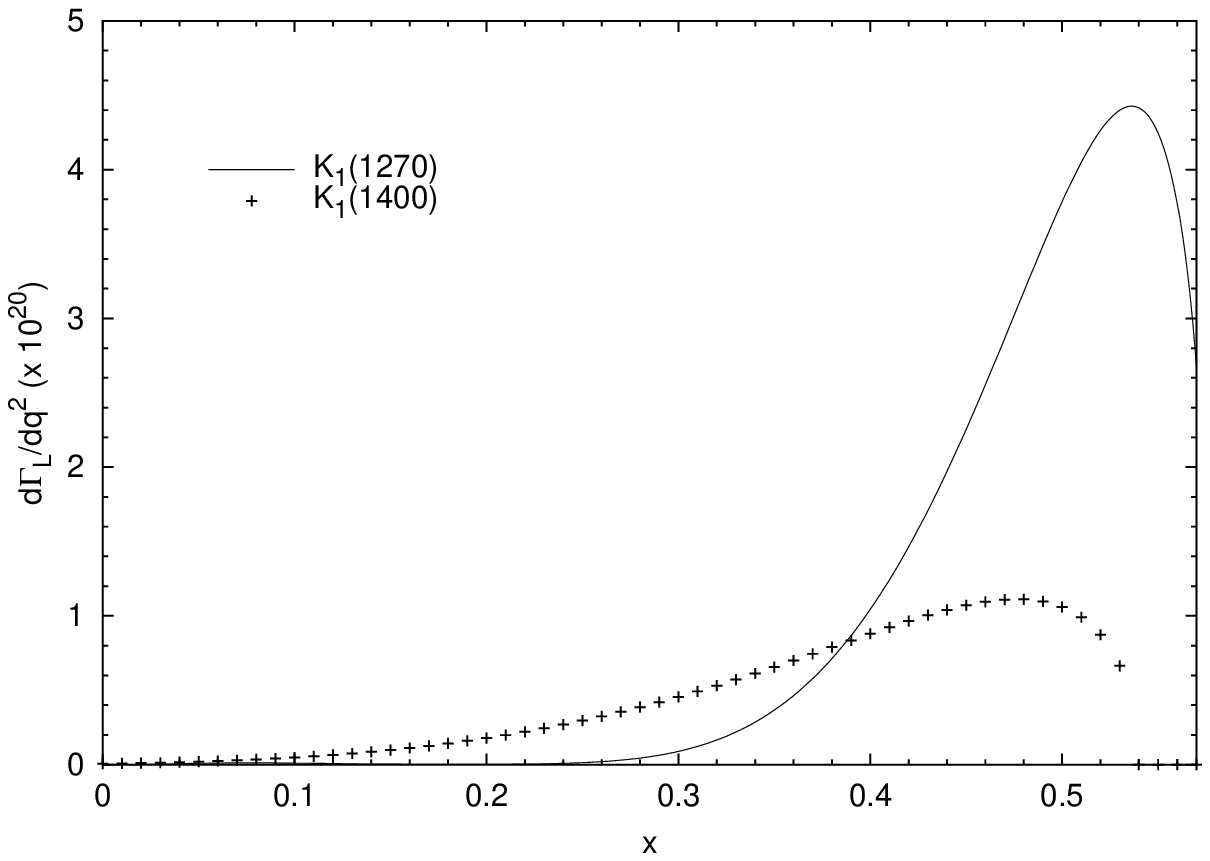}
\end{center}
\caption{The dependence of the $d\Gamma_{L}/dq^{2}$ on
$x=q^{2}/m_{B}^{2}$}.\label{fig1}
\end{figure}

\begin{figure}
\vspace*{-1cm}
\begin{center}
\includegraphics[width=10cm]{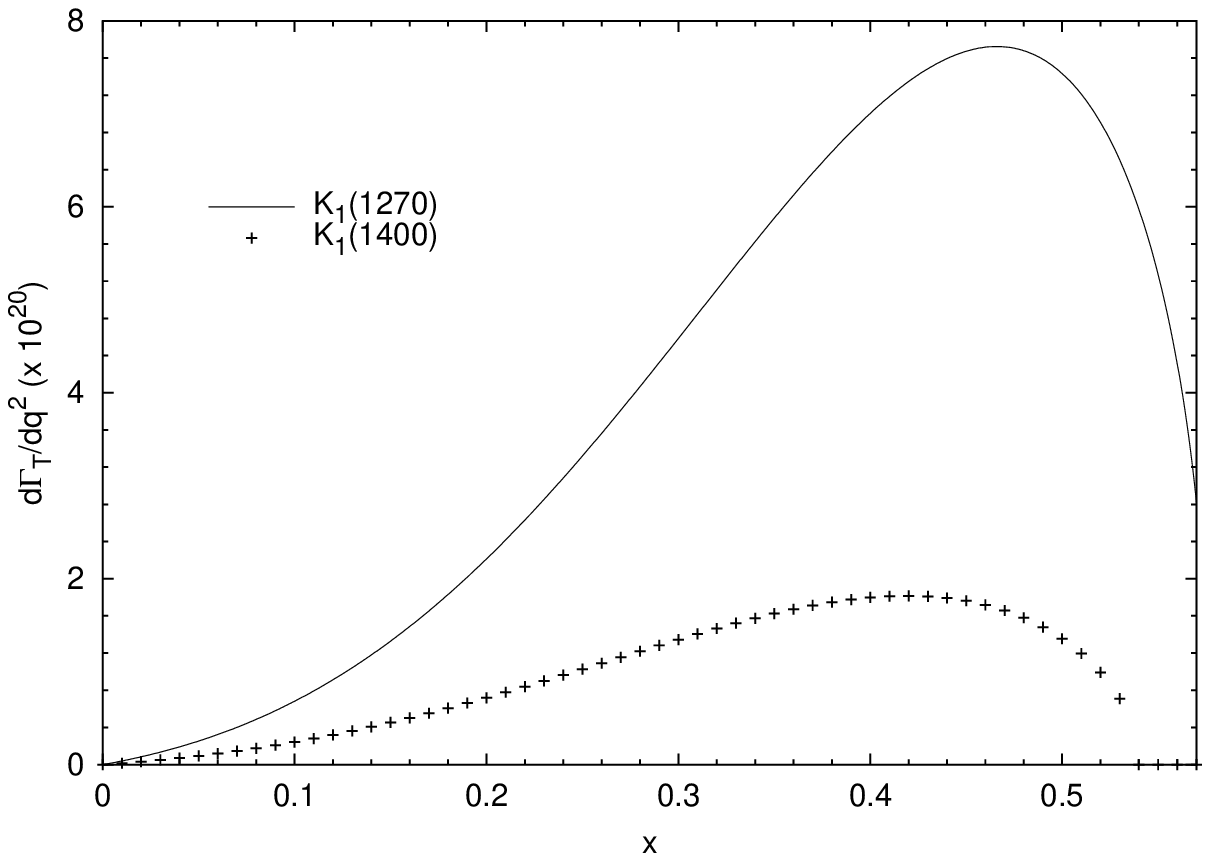}
\end{center}
\caption{The dependence of the $d\Gamma_{T}/dq^{2}$ on
$x=q^{2}/m_{B}^{2}$}.\label{fig2}
\end{figure}

 Having the $q^2$ dependent expressions for the form factors, in the
 following part, we present the evaluation of the numerical values
 for the longitudinal and transverse components of the differential
 decay width, branching ratio and the asymmetry parameter of $K_{1}$
 meson. Figures \ref{fig1} and \ref{fig2} depict the dependency of the
 $d\Gamma_{L}/dq^{2}$ and $d\Gamma_{T}/dq^{2}$ on $x=q^{2}/m_{B}^{2}$
 in the allowed physical region for $x$, i.e., $0<x<0.57$ for $K_{1}(1270,1400)$. From these
 figures, we see that the longitudinal and transverse  components of the decay rate
 for $K_{1}(1270)$ is about four times greater than that of the  $K_{1}(1400)$.
 The  maximum values
 for the $d\Gamma_{L}/dq^{2}$ is at $x=0.48$ and $x=0.52$
 for $K_{1}(1400)$ and $K_{1}(1270)$, respectively. However, the peak values for
   these states are at
    $x=0.43$ and
    $x=0.46$ for the $d\Gamma_{T}/dq^{2}$ case. These figures also show that the
     transverse component of the differential decay
    rates have nearly the same behavior  at low   values of the
    $q^2$, but in this region, the $d\Gamma_{L}/dq^{2}$ for $K_{1}(1270)$ is  zero and
    has almost nonzero value for $K_{1}(1400)$. Finally, we denote the
    dependency of the asymmetry parameter $\alpha$,
 characterizing the polarization of the $K_{1}$ meson in
 terms of the $x$. This figure depict that for both $K_{1}(1270)$ and $K_{1}(1400)$ cases, the
 $\alpha$ varies in the interval $-1<\alpha<1$ in the allowed region of the $q^2$.
 For $K_{1}(1270)$, the asymmetry parameter is unity with negative sign in the
 interval  $0<x<0.3$. This implies that, in this region, the longitudinal
 component of the differential decay rate is zero and this is in agreement with
 Fig. 1. After $x=0.3$, the $d\Gamma_{L}/dq^{2}$ approaches to zero and changes sign at $x=0.5$.
  In this point, the longitudinal component of the decay rate is near
  to  its
  maximum and half of the transverse component.  The $\alpha$ for  $K_{1}(1270)$
   has positive sign in the region $0.50<x<0.57$. The asymmetry parameter for
   $K_{1}(1400)$ changes sign at $x=0.4$ and has a minimum at about
   $x=0.1$. Any measurement on the asymmetry parameter and
   determination of its sign can give valuable information about the
   structures of the $\mid K_{1}^{A}>$ and $\mid K_{1}^{B}>$ states.
\begin{table}[h]
\centering
\begin{tabular}{|c||c|} \hline
 $\textit{\textbf{B}}(B\rightarrow
K_{1}(1270)\nu\overline{\nu})$ & $\textit{\textbf{B}}(B\rightarrow
K_{1}(1400)\nu\overline{\nu})$
\\\cline{1-2}\hline\hline
  $(3.54\pm1.27)\times10^{-6}$ &
$(8.92\pm3.21)\times10^{-7}$
\\\cline{1-2}
\end{tabular}
\vspace{0.8cm} \caption{Values for the branching ratio of
$B\rightarrow K_{1}\overline{\nu}\nu$.} \label{tab:3}
\end{table}

 Taking into account the $q^2$ dependency of the form
factors and performing integration over $q^2$ for
$d\Gamma_{tot}/dq^{2}$ (in Eq. \ref
 {total}) in the interval $0\leq
q^2\leq(m_{B}-m_{K_{1}})^2$ and using the total life-time
$\tau_{B}=1.638\times10^{-12}s$ \cite{Yao}, the branching ratio for
$B\rightarrow K_{1}(1270, 1400)\nu\overline{\nu}$ is obtained as
Table 3. This Table shows that the branching fraction for
$K_{1}(1270)$ about four times greater than that of the
$K_{1}(1400)$ states.

 In conclusion, separating the mixture of the $ K_{1}(1270)$ and $K_{1}(1400) $
states, the form factors related to the $B\rightarrow K_{1}(1270,
1400)\overline{\nu}\nu$
 decay were calculated using
three-point QCD sum rules approach. Taking into account the $q^{2}$
dependencies of the form factors, the longitudinal and transverse
component of the differential decay width as well as the asymmetry
parameter $\alpha$, characterizing the polarization of the $K_{1}$
meson and the branching ratio of these transitions were evaluated.

\section{Acknowledgment}
  The authors would like to thank T. M. Aliev and A. Ozpineci for
  their useful discussions and also TUBITAK, Turkish Scientific and Research
Council, for their financial support provided under the project
103T666.

\clearpage

\begin{figure}
\vspace{-2cm}
\begin{center}
\includegraphics[width=10cm]{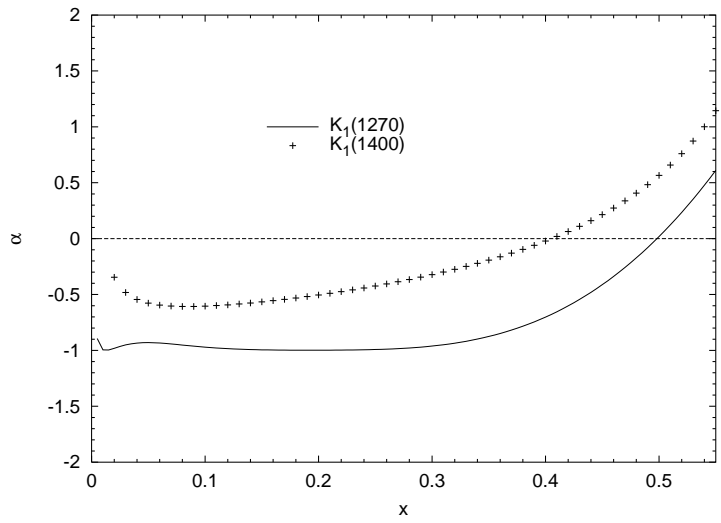}
\end{center}
\caption{The dependence of the asymmetry parameter $\alpha$ on
$x=q^{2}/m_{B}^{2}$.} \label{fig9}
\end{figure}

\end{document}